\documentclass[%
reprint,
amsmath,amssymb,
aps,
pra,
]{revtex4-1}

\usepackage{color}
\usepackage{mathptmx}
\usepackage{geometry}
\usepackage{ulem}
\usepackage{bm}
\usepackage[pdftex,unicode,colorlinks]{hyperref}
\usepackage{epstopdf}
\usepackage{graphicx}
\usepackage{dcolumn}
\usepackage{bm}
\usepackage{hyperref}
\usepackage{ulem}

\newcommand{\mean}[1]{\langle#1\rangle}

\newcommand{\partd}[2]{\displaystyle\frac{\partial#1}{\partial#2}}

\DeclareMathOperator{\asinh}{asinh}

\begin{document}

\title{Detection loss tolerant supersensitive phase measurement with an SU(1,1) interferometer}

\author{Mathieu Manceau$^{1}$, Gerd Leuchs$^{1,2}$, Farid Khalili$^{3,4}$ and Maria Chekhova$^{1,2,3}$}

\affiliation{$^1$Max-Planck-Institute for the Science of Light, Erlangen, Germany\\
$^2$University of Erlangen-N\"urnberg, Staudtstrasse 7/B2, 91058 Erlangen, Germany\\
$^3$Faculty of Physics, M V Lomonosov Moscow State University, Moscow, Russia\\
$^4$Russian Quantum Center, Skolkovo 143025, Russia}

\date{\today}
\begin{abstract}
In an unseeded SU(1,1) interferometer composed of two cascaded degenerate parametric amplifiers, with direct detection at the output, we demonstrate
a phase sensitivity overcoming the shot noise limit by $2.3$~dB. The interferometer is strongly unbalanced, with the parametric gain of the second
amplifier exceeding the gain of the first one by a factor of 2, which makes the scheme extremely tolerant to detection losses. We show that by
increasing the gain of the second amplifier, the phase supersensitivity of the interferometer can be preserved even with detection losses as high
as $80$\%. This finding can considerably improve the state-of-the-art interferometry, enable sub-shot-noise phase sensitivity in spectral ranges with
inefficient detection, and allow extension to quantum imaging.
\end{abstract}


\maketitle

The sensitivity of an interferometric measurement on a phase shift depends on the state of light used as a probe and the measurement scheme. 
A `standard' precision is provided by a coherent state fed into a Mach-Zender interferometer, the so-called shot noise limit (SNL). A 
measurement beating this limit is said to be supersensitive. In order to make super-sensitive phase measurements, quantum resources can be used. 
First proposed~\cite{Caves1981} and experimentally tested~\cite{Xiao1987,Grangier1987} in the 1980-s, squeezed light is now used for gravitational 
wave detection~\cite{LIGO_2011,Aasi2013} beyond the shot noise limit. Squeezed states can improve the sensitivity in the presence of 
loss~\cite{Gea1987}, a finite interference visibility~\cite{Gea1989} and their use is compatible with power recycling~\cite{Brillet1991}. 
Supersensitivity can also be achieved with other quantum states~\cite{Holland1993,Lee2002}, which, however, are difficult to produce.

Besides the input state and the detection scheme, one can also modify the interferometer. Yurke et al.~\cite{Yurke1986} proposed to
use cascaded optical parametric amplifiers (OPAs) instead of the passive beam-splitters of conventional interferometric setups, the phase sensitive
response of the OPAs giving rise to interference patterns. Such interferometers, usually called SU(1,1) interferometers, can display phase
super-sensitivity without seeding the amplifiers~\cite{Yurke1986}. Seeding can be used in order to increase the number of sensing photons and
therefore the overall sensitivity~\cite{Plick2010}. It was recently  noted theoretically that such a scheme involving two amplifiers can help
overcoming the deleterious effects of optical losses~\cite{Ou2012,Marino2012} on phase sensitivity. Note that losses occurring inside the
interferometer have a different impact on the phase sensitivity than losses outside of the interferometer. These two kinds of
losses are therefore distinguished in the following and respectively called internal and external/detection losses. The influence of the latter kind
of losses can be suppressed by the second amplifier. Indeed, amplifying the signal with the second OPA at the output of the interferometer while
keeping the probing field constant eventually eliminates the effect of detection losses~\cite{Sparaciari2016,Manceau2017}.

Optical SU(1,1) interferometers have been implemented using two cascaded four-wave mixers~\cite{Jing2011,Kong2013}. Recent experiments with this
interferometer operating in the continuous-wave regime with seeding have demonstrated phase super-sensitivity~\cite{Hudelist2014} using homodyne
detection.  A truncated version, \textit{i.e.} without a second amplifier, of the same scheme~\cite{Anderson2016} has also demonstrated the
possibility of supersensitive phase measurement. The effect of internal losses on the quantum noise of an SU(1,1) interferometer has been studied in 
detail in Ref.~\cite{Xin2016}. Finally, an atom SU(1,1) interferometer has shown sensitivity beyond the SNL~\cite{Linnemann2016} and more complex 
schemes based on SU(1,1) interferometry such as an atom-light hybrid interferometer~\cite{Chen2015} or a so called pumped-up SU(1,1) 
interferometer~\cite{Szigeti2017} are also currently investigated.

\begin{figure} \centering \includegraphics[scale=0.25]{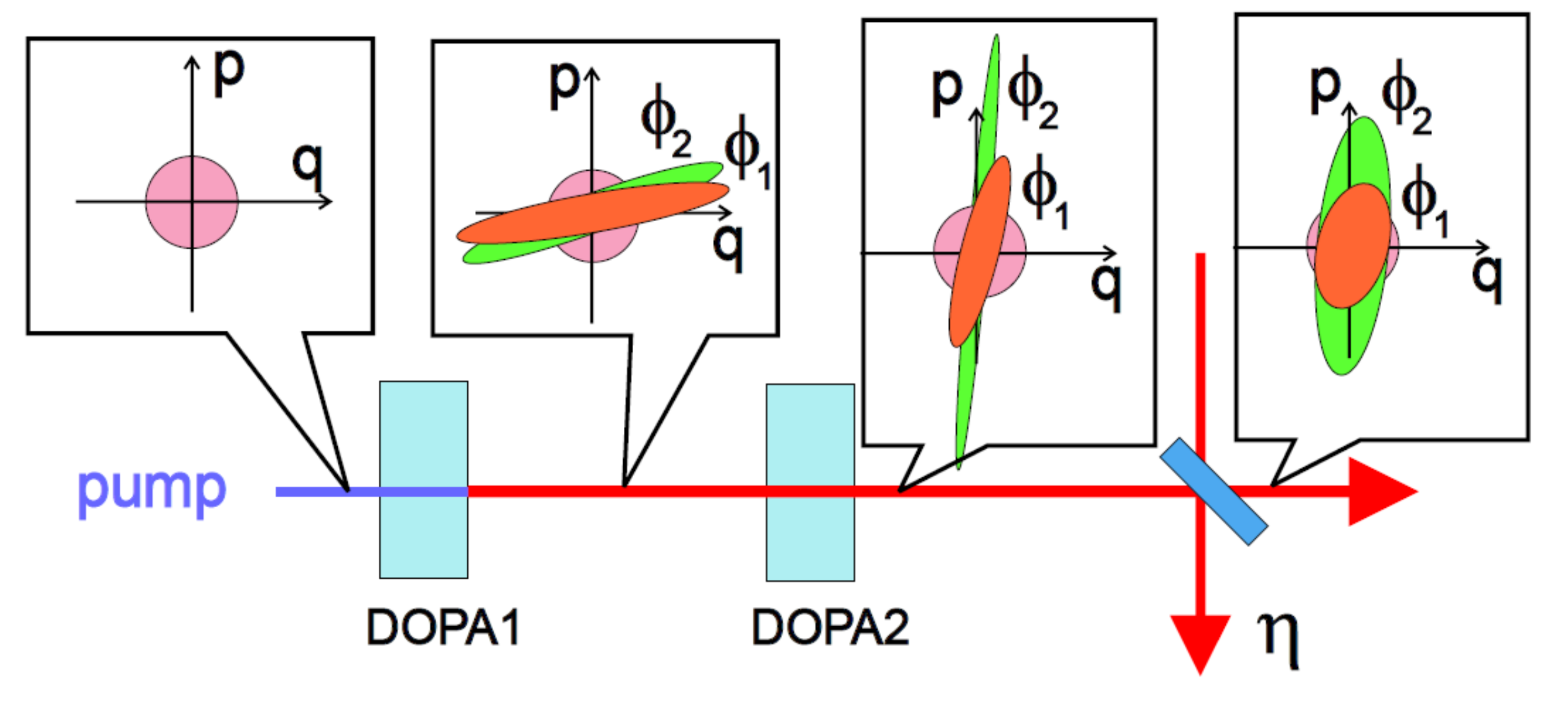} \caption{Unbalanced unseeded SU(1,1) interferometer formed by coherently
pumped DOPA1 and DOPA2. Squeezed vacua with phases $\phi_{1,2}$ are distinguishable if their Wigner functions do not fully overlap. The squeezing
after DOPA2 depends on the phase $\phi_{1,2}$ but the distinguishability remains the same. A low detection efficiency $\eta$ almost destroys the
squeezing but, with the DOPA2 gain high enough, the states remain distinguishable.} \label{fig:Explanation} \end{figure}

In this letter we report phase supersensitivity in an SU(1,1) interferometer formed by two cascaded nonlinear $\chi^{(2)}$ crystals. Our goal is to
overcome the negative effect of detection losses by pumping the second OPA stronger than the first one. In the previous realizations of the SU(1,1)
interferometer, both OPAs had the same absolute values of the parametric gain. However as shown theoretically in Ref.~\cite{Manceau2017}, the
gain unbalancing leads to an increased tolerance of the setup to the detection losses. The idea is illustrated in Fig.~\ref{fig:Explanation}.

The interferometer consists of two degenerate parametric amplifiers, DOPA1 and DOPA2, the first one transforming an input vacuum state into a squeezed
vacuum state, which then acquires a phase $\phi$ to be measured. Two squeezed vacuum states differing by their phases $\phi_1$ and
$\phi_2=\phi_1+\Delta\phi$ can be distinguished whenever their Wigner functions (shown by orange and green colors) do not completely overlap. Such
squeezed vacua have a higher distinguishability than two coherent states with the same photon number and the same phase difference $\Delta\phi$. The
second amplifier DOPA2 applies squeezing opposite to the one of DOPA1, and the resulting output states are now squeezed in other directions, and
differently as long as $\phi_1\ne\phi_2$. Importantly, the squeezing operation does not change their distinguishability. In the presence of low
detection efficiency $\eta$, represented in Fig.~\ref{fig:Explanation} by a beamsplitter, the states change dramatically, so that the noise in the
squeezed quadrature becomes almost equal to the shot noise. At the same time, the spread of the anti-squeezed quadrature is only reduced by a factor
$\sqrt{\eta}$, and will remain different for the two states as long as it is initially large. Therefore, provided that the DOPA2 gain is high enough,
and the states at its output are highly squeezed, they are still distinguishable after the losses are applied. Note that the smaller the detection
efficiency $\eta$, the stronger squeezing in DOPA2 is required to reach a certain phase sensitivity \cite{Manceau2017}.

Although the same principle of unbalancing can be applied to various configurations of an SU(1,1) interferometer (seeded or unseeded, with direct or
homodyne detection) and even to an SU(2) interferometer fed with squeezed light and followed by another squeezer~\cite{Manceau2017}, here we use the
simplest configuration of an unseeded interferometer with direct intensity detection. The robustness of the measured phase sensitivity against
detection losses is tested for various parametric gains of the DOPAs. We achieve a high parametric gain by using intense picosecond pump pulses and
show the possibility to overcome the detrimental effect of losses for large parametric gain of the second DOPA.

\begin{figure} \centering \includegraphics[scale=0.70]{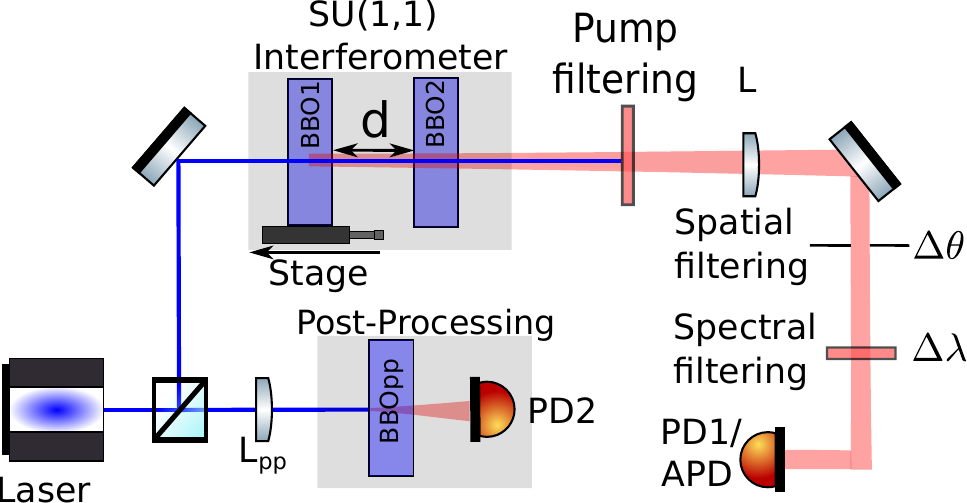} \caption{ The unseeded SU(1,1) interferometer is made of two cascaded BBO crystals,
BBO1 and BBO2. The generated PDC light is spatially filtered by a pinhole placed in the focal plane of the lens L and spectrally filtered by an
interference filter. The number of photons contained in a given PDC pulse is measured by the photodiode PD1. A second photodiode PD2 is used for
post-selecting a pump intensity interval. For finding the SNL, BBO2 is removed and an APD registers the number of photons emitted by BBO1 within the 
$\Delta\lambda$, $\Delta\theta$ bandwidths.} \label{fig:Setup} \end{figure}

The setup is depicted in Fig.~\ref{fig:Setup}. The SU(1,1) interferometer consists of two cascaded $3$~mm long $\beta$-Barium Borate (BBO) crystals
cut for collinear frequency-degenerate type-I phase matching. The pump is the second harmonic of a Spectra Physics Spitfire Ace system with a $5$~kHz
repetition rate, $1.5$~ps pulses and a central wavelength of $\lambda_p = 400$~nm. The beam diameter is $0.2$~mm full width at half maximum
(FWHM) inside the SU(1,1) interferometer. Two dichroic mirrors are subsequently used to eliminate the pump beam. The parametric down converted (PDC)
light generated by the SU(1,1) interferometer is spatially filtered by a pinhole in the focal plane of a $f=90$~cm focal length lens selecting an
angular bandwidth $\Delta \theta$. An interference filter further selects a spectral bandwidth $\Delta\lambda$ around the central wavelength
$\lambda_s = 800$~nm. After the spatial and spectral filtering, the PDC light is detected using a low-noise charge integrating p-i-n diode
(PD)\cite{Hansen2001,Iskhakov2009,Agafonov2010_2} (see supplementary).

The first crystal BBO1 is fixed on a translation stage. The distance $d$ between the two crystals can therefore be changed. Due to the dispersion of
the air~\cite{Perez2014}, the squeezed light and the pump acquire different phase shifts $\phi_s\ne\phi_p$ in the air gap, and the
phase $\phi=\phi_s-\phi_p/2$ of the squeezed state at the input of DOPA2 (Fig.~\ref{fig:Setup}) can be scanned by adjusting the distance $d$. A set
of $4000$ post-processed pulses is measured for each position of BBO1. To avoid high uncertainty in the measured mean photon
number of the interferometer output light, the pump intensity fluctuations are eliminated through pulse post-selection (see supplementary).

By measuring the transmission of the setup at $\lambda_s = 800$~nm, we found that the internal losses, caused by reflections on a single facet of each
BBO crystal, are $3\%$. The external transmission of the setup, including the detector quantum efficiency, is measured to be $\eta=77\%$. It
corresponds to the losses for a set of plane waves within the $\Delta\lambda$, $\Delta\theta$ bandwidths of the frequency and spatial filters.
Assuming Gaussian noise and sufficiently small phase fluctuations, the phase sensitivity $\Delta\phi$ is given by
\begin{equation}\label{eq:DeltaPhiDef}
    \Delta\phi
    = \frac{\Delta N}{\left|\partd{\mean{N_f}}{\phi}\right|}
  \,,
\end{equation}
$\Delta N$ including the intrinsic photon noise of the measured light and the detector noise and $\mean{N_f}$ being the mean number
of photons at the interferometer output. For a given position $d$ of BBO1, both values are determined from the set of post-processed pulses. The
phase sensitivity is then calculated according to Eq.~\ref{eq:DeltaPhiDef}.

The distance between the two crystals is converted into a phase knowing the periodicity of the interference (see supplementary). Figure
\ref{fig:1stFigure} presents typical phase sensitivity measurements. The shot noise limited phase sensitivity for a set of plane waves within the
$\Delta\lambda$, $\Delta \theta$ bandwidths, $\Delta\phi_{SNL}$, is defined as
\begin{equation}\label{eq:DeltaPhiSNL}
  \Delta\phi_{SNL}
  = \frac{1}{2\sqrt{\langle N(\Delta\lambda,\Delta\theta)\rangle}},
\end{equation}
with $\mean{N(\Delta\lambda,\Delta \theta)}$ being the mean number of photons inside the interferometer within the $\Delta\lambda$, $\Delta\theta$
bandwidths~\cite{Manceau2017}. It is measured by removing BBO2 and registering the number of photons within the $\Delta\lambda$, $\Delta\theta$
bandwidths with an avalanche photodiode (APD) after attenuating the beam with neutral density filters. The value of 
$\mean{N(\Delta\lambda,\Delta \theta)}$ is found from the APD count rate knowing the APD detection efficiency $\eta_{APD}=47\%$ and the filters 
transmission. Note that the resulting number of photons is considerably less than the number of photons in a single Schmidt mode 
$N_{mode}=\sinh(r_1)^2$ because of the strong spatial filtering. For a single Schmidt mode of the multimode OPA, this narrowband detection 
leads to a reduction of quantum efficiency~\cite{Rytikov2008,Agafonov2010_2,Perez2015} from $\eta$ to $\eta_{SM}=\nu\eta$. Here, 
$\nu$ corresponds to the transmission of the spatial and frequency filters for the Schmidt mode under consideration. 

The SNL phase sensitivity associated with a single Schmidt mode, $\Delta\phi_{SM}=\frac{1}{2 \sinh(r_1)}$,
can still be beaten in theory with our large values of $|r_2|$~\cite{Sparaciari2016,Manceau2017}. However, the non ideal visibility, here being
$\mathcal{V}=97\%$ for the given spatial and frequency filtering (see supplementary), is a limiting factor for the best phase sensitivity achievable
\cite{Gea1989}. Indeed, as the working point of this interferometer is close to the dark fringe, the non-unity visibility acts as an extra source of 
noise. $\Delta\phi_{SM}$ can therefore not be reached with the current experimental configuration. At the
same time, as the SNL is defined for coherent light, it is still determined by the number of photons inside the interferometer within the chosen
bandwidths, $\mean{N(\Delta\lambda,\Delta \theta)}$, regardless of the Schmidt mode structure.

Fig.~\ref{fig:1stFigure}a shows two measurements with different parametric gain values $r_1$ of the first crystal, and therefore different mean 
numbers of photons $\mean{N(\Delta\lambda,\Delta \theta)}$, varied by changing the phase matching. The blue points are for 
$\mean{N(\Delta\lambda,\Delta\theta)}=1.7$ photons and $r_1=1.5$. The corresponding shot noise limited phase sensitivity is 
$\Delta\phi_{SNL}=0.38$~rad (dashed blue line, the uncertainty shown by the blue rectangle). The best phase sensitivity is $2.3$~dB below the SNL. 
For 
a larger number of photons in the interferometer, $\mean{N(\Delta\lambda,\Delta \theta)}=11$, corresponding to $r_1=2.7$ and the gain of the second 
crystal almost the same, the SNL is overcome by $1.2$~dB (red points). The inset shows the $\langle N_f(\phi)\rangle$ dependences.
\begin{figure}
  \centering
\includegraphics[scale=0.5]{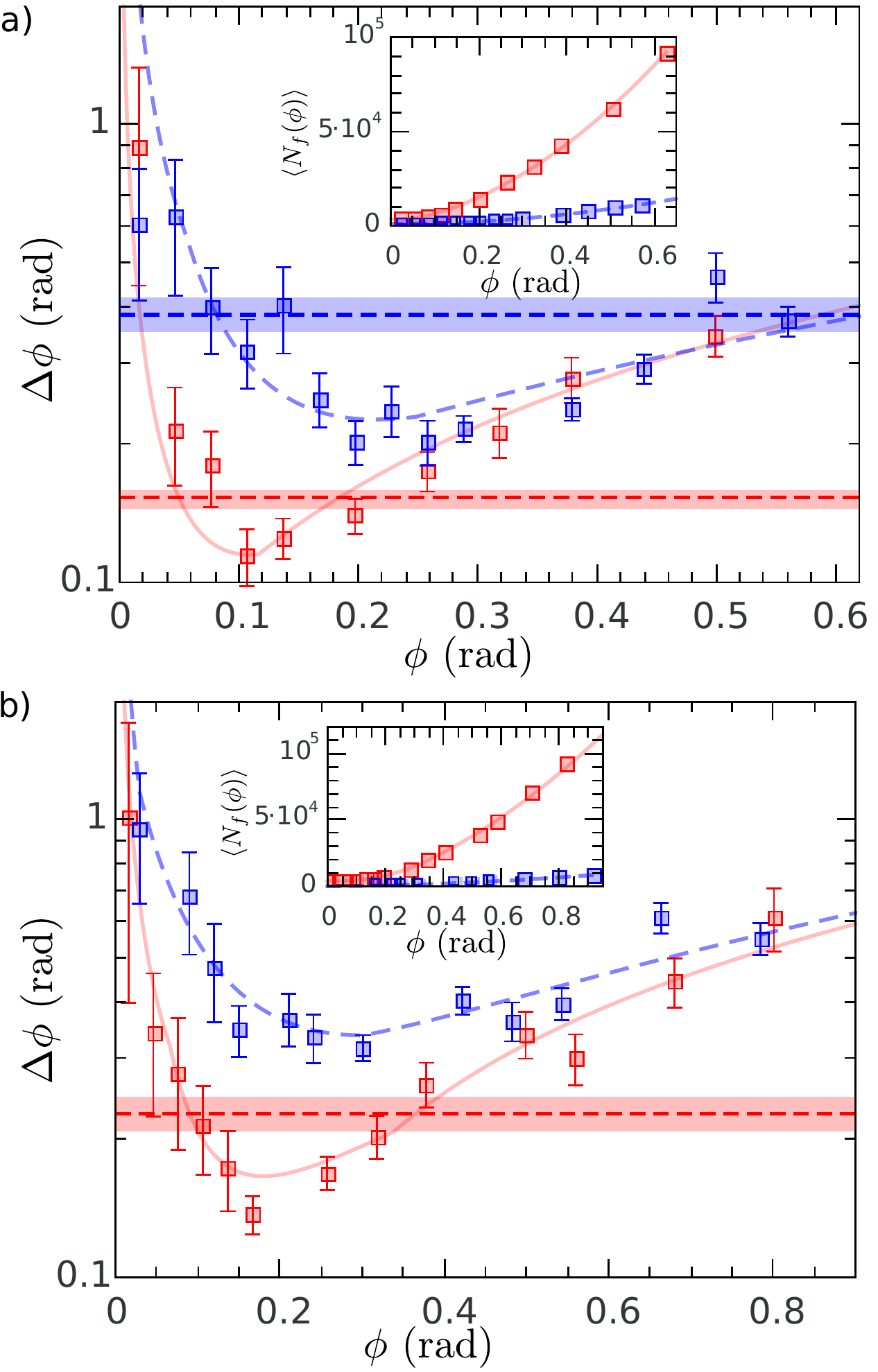}
  \caption{Phase sensitivity measurements with $\Delta\theta=0.9$~mrad and $\Delta\lambda=3.1$~nm. a) Phase sensitivity for strongly differing gain
values of BBO1: $r_1=1.5$,  $\mean{N(\Delta\lambda,\Delta \theta)}=1.7$~photons (blue), and  $r_1=2.7$,
$\mean{N(\Delta\lambda,\Delta\theta)}=11$~photons (red). The values of $|r_2|$ are $4.9$ and $5.2$ respectively. The dashed lines mark the SNL
$\Delta\phi_{SNL}$ for the two cases. b) Phase sensitivity for strongly differing gain values of BBO2:  $|r_2|=3.9$ (blue),
$|r_2|=5.2$ (red). The values of $r_1$ are $2.5$ and $2.1$ respectively. $\mean{N(\Delta\lambda,\Delta \theta)}=4.8$~photons is the same in both
cases and the corresponding $\Delta\phi_{SNL}$ is given by the dashed red line. The insets show the $\langle N_f(\phi)\rangle$ dependences.}
  \label{fig:1stFigure}
\end{figure}

Fig.~\ref{fig:1stFigure}b shows that a larger parametric gain of the second crystal leads to an improved sensitivity. Indeed, for $|r_2|=3.9$ (blue 
points), the best phase sensitivity is $\Delta\phi_{min}=0.33$~rad while a larger parametric gain $|r_2|=5.2$ obtained by 
increasing the pump power gives $\Delta\phi_{min}=0.16$~rad (red points). The number $\mean{N(\Delta\lambda,\Delta \theta)}=4.8$ is kept constant by 
controlling the phase matching of BBO1. Hence the shot noise limit is the same in both cases and it is not beaten for the lower gain case.

\begin{figure}
  \centering
\includegraphics[scale=0.55]{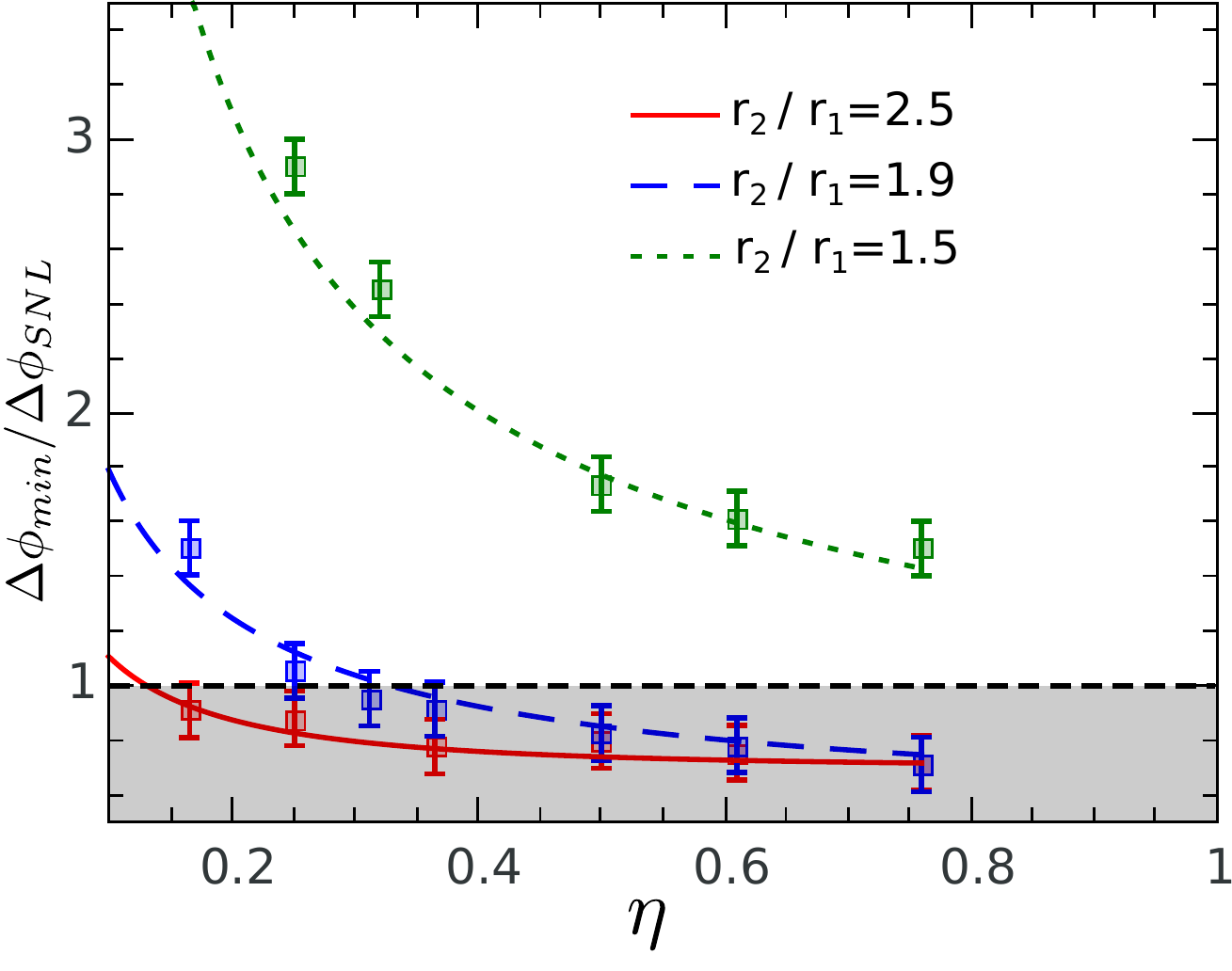}
  \caption{Best phase sensitivity $\Delta\phi_{min}$ normalized to $\Delta\phi_{SNL}$ against the detection
transmission $\eta$, reduced by adding neutral density filters. The number of photons is kept constant for all presented cases, with
$\langle N(\Delta\lambda,\Delta\theta)\rangle=4.5\pm0.5 $. Red: $r_1=2.1$, $|r_2|=5.2$. Blue: $r_1=2.5$,
$|r_2|=4.7$. Green: $r_1=2.6$, $|r_2|=4$.}
  \label{fig:2ndFigure}
\end{figure}

Finally, we test the tolerance of the scheme to detection losses. For a given set ($r_1,|r_2|$) of parametric gains, we vary the
loss by adding neutral density filters before the detection. Figure \ref{fig:2ndFigure} shows the measurement results for three sets of
parametric gains. Theoretical dependences taking into account the detector noise and the non-ideal visibility are also shown (see
supplementary). The number of photons is kept constant, $N(\Delta\lambda,\Delta\theta)=4.5\pm0.5$, for all three cases. The parametric gain
of the second crystal is reduced from $|r_2|=5.2$ (red points) to $|r_2|=4.7$ (blue points) and $|r_2|=4$ (green points). We can observe
that for a large unbalancing of the interferometer, $|r_2|/r_1=2.5$ (red curve), the phase sensitivity is robust against the added losses,
the system being still supersensitive for $\eta=17$\%. For a smaller unbalancing of $|r_2|/r_1=1.9$ (blue curve), the sensitivity is
degrading faster with the added losses. Finally, if the unbalancing is too weak $|r_2|/r_1=1.5$ (green curve), as was the case already in
\ref{fig:1stFigure}b, even without adding extra losses the interferometer cannot compensate for the detector noise and the sensitivity is
therefore always above the SNL.

We would like to stress that this active detection strategy involving parametric amplification can also be implemented at the output of more common 
SU(2) interferometers~\cite{Manceau2017} (Mach-Zender, Michelson, Sagnac etc.). This is especially important in gravitational wave detectors where the 
detection losses considerably limit  the sensitivity. Indeed, external elements of interferometers such as the Faraday isolators, output mode cleaners 
and photodetectors are reported as the  main sources of losses \cite{LIGO_2011}. Also the tolerance of the interferometer to detection losses is very 
important in experiments where a high phase sensitivity is required at frequency ranges (infrared, terahertz) where the efficiency of detectors is low 
for technical reasons. In addition, the unbalanced SU(1,1) interferometer can be reconfigured to be applied to sub-shot-noise imaging~\cite{Brida2010} 
where the detection efficiency plays a crucial role.

In conclusion, we have demonstrated for the first time phase supersensitivity in an unseeded SU(1,1) interferometer, beating the shot noise limit by 
$2.3$ dB. Moreover we have shown that a gain-unbalanced interferometer is tolerant to external losses. With the parametric gain $5.2$, the SNL was 
beaten even for a detection efficiency as low as $17\%$. This result is relevant to many cases where detection losses are high, such 
as gravitational-wave detection and phase measurements in the infrared and terahertz ranges.

We acknowledge the financial support of the joint DFG-RFBR project CH1591/2-1 - 16-52-12031 NNIOa

%

\appendix

\section{Detection}

\begin{figure}[h] \centering \includegraphics[scale=0.5]{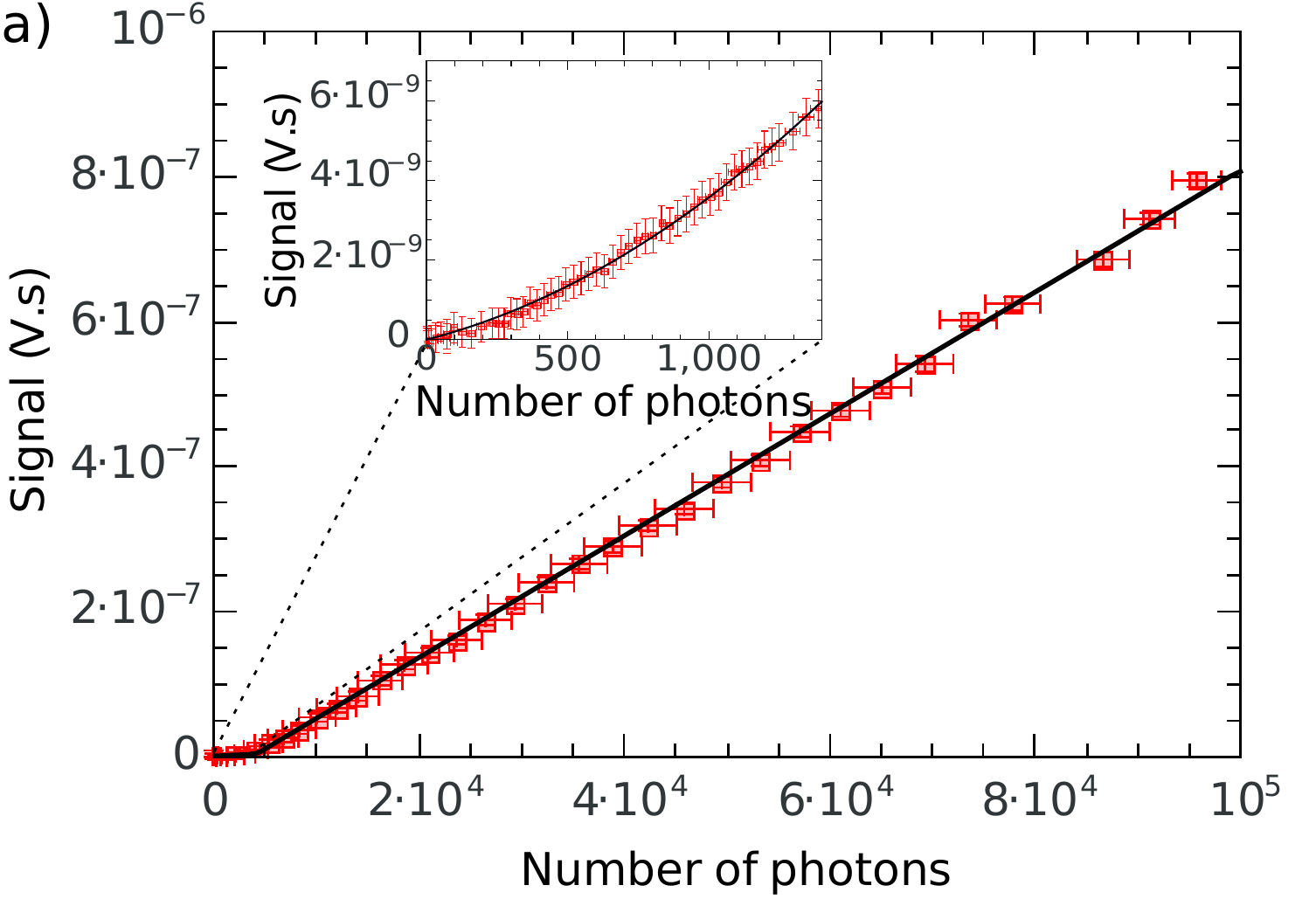} \caption{a) Detector calibration curve, the mean area of 
the output voltage pulse versus the incoming mean number of photons per pulse. Above the level of $2000$ photons per pulse, the 
detector shows a linear response. Inset: Zoom into the low-photon number part of the calibration curve showing a nonlinear behavior 
below $2000$ photons per pulse.} \label{fig:Calibration} \end{figure}

The detectors are Hamamatsu $S3883$ p-i-n diodes with quantum efficiency $90\%$ at $800$~nm, followed by pulsed charge-sensitive amplifiers based on 
Amptek A250 and A275 chips, with peaking time $2.77$~$\mu$s. For each input light pulse, the output voltage pulse has a 8~$\mu$s duration and the 
area 
depending on the input number of photons. The fundamental beam of the Spectra Physics Spitfire Ace system at $\lambda = 800$~nm is used to calibrate 
the detector. Figure \ref{fig:Calibration}a shows the mean area of the output voltage pulse for a given mean number of incoming photons per pulse.  
For each set of measurements presented in this work, the mean number of photons is derived from this calibration curve. One can see in 
Fig.~\ref{fig:Calibration}a that the response of the detector to the light pulses is linear within most of the photon-number range under study in 
this 
work. However, as it is apparent in the inset of Fig.~\ref{fig:Calibration}a for low photon fluxes ($\lesssim 2000$ photons per pulse), a 
nonlinearity, with a corresponding decrease in the detection sensitivity of the detector, needs to be taken into account.

Also, the detector brings some additional noise to the measurements. The corresponding noise in terms of photons per pulse is $\Delta 
N_d=290$ photons in the linear part of the curve and considerably larger in its nonlinear part. The measurements presented in the main text are 
characterized by mean numbers of photons within a range of a few hundreds to a thousand in the region of interest for supersensitivity, \textit{i.e.} 
around the dark fringe. A detector noise of $\Delta N_d\simeq1000$ photons per pulse needs to be taken into account in this case.


\section{Period of interferences and visibility}

\noindent The interference period $D$ is~\cite{Perez2014}

\begin{equation}\label{eq:Period}
D=\frac{2\lambda_p}{\delta n_{air}},
\end{equation}

\noindent with $\lambda_p$ the pump wavelength and $\delta n_{air}=n_{air}(400nm)-n_{air}(800nm)$ the dispersion of the air gap between the two 
crystals for the considered pump and signal wavelengths. The laboratory temperature and humidity conditions give $D=52$~mm. The phase $\phi$ can 
therefore be determined from $D$ and the relative distance between the two crystals.  A phase shift $\phi=\pi$ is equivalent to a $52$~mm distance 
between the two crystals.

\begin{figure}[h]
  \centering
\includegraphics[scale=0.5]{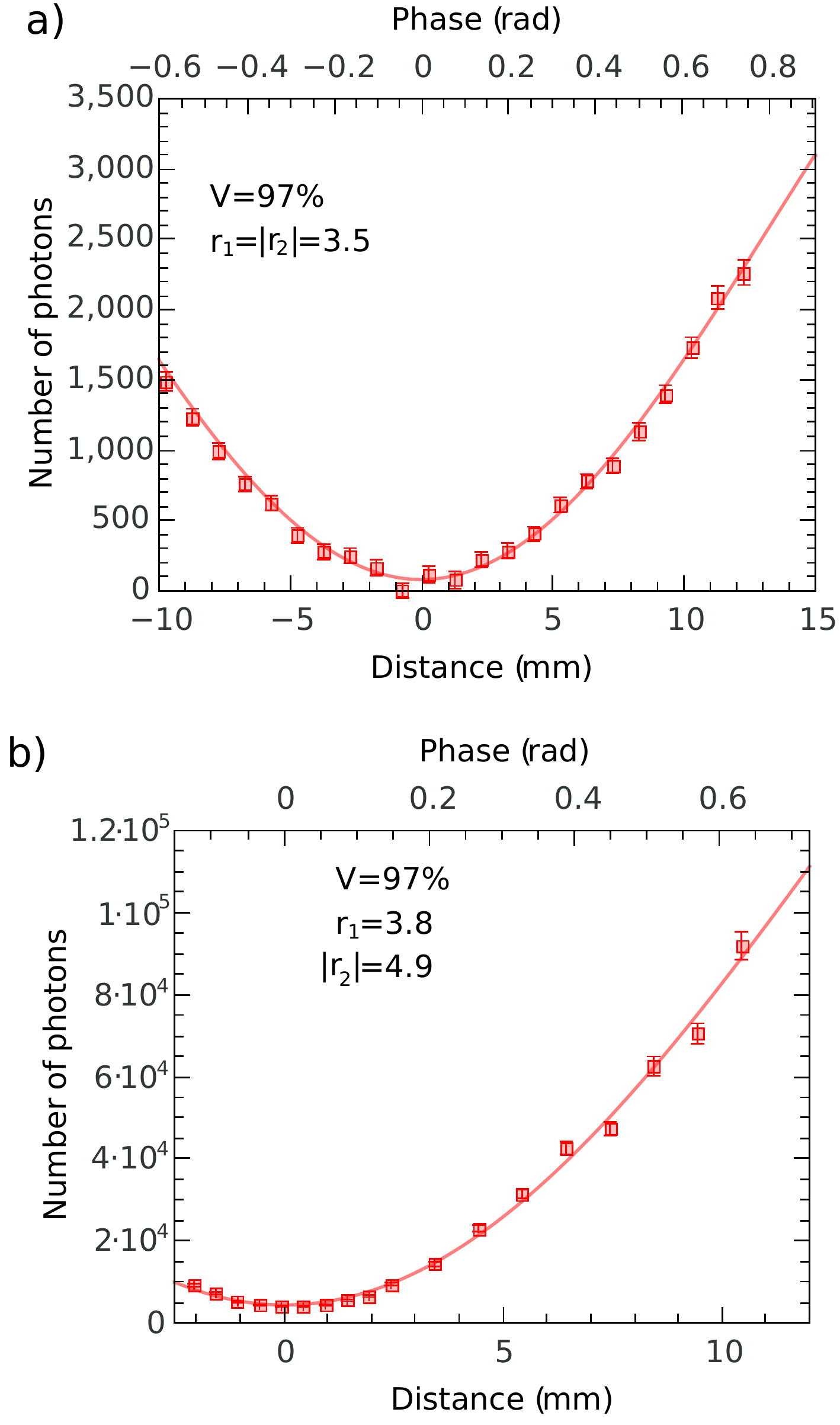}
  \caption{Typical fringe patterns obtained for two sets of measurements characterized by different gain values. The fit is made
with the expected $D=52$~mm period from which the phase $\phi$ can be calculated.}
  \label{fig:Visibility}
\end{figure}

Figure \ref{fig:Visibility} shows typical fringe patterns for two sets of parametric gain values. The distance between the crystals and the phase are 
arbitrarily set such that $\phi=0$ corresponds to the dark fringe. A fit with  Eq.~(\ref{N2}) gives a visibility $\mathcal{V}=97\%$ in both cases, as 
well as for all the measurements reported in this article. The main reason for the visibility not exceeding $97$\% is that the interference is 
observed for spatially separated crystals. In this case, emission from both of them is partially distinguishable and a $100$\% visibility can be 
achieved only by selecting a single plane wave, which would be the case for an infinitely small pinhole, $\Delta\theta=0$.

\section{Curve fitting, gain values} 

Following the calculations presented in Ref.~\cite{Manceau2017}, the interferometer output mean number of photons $\mean{N_f}$ is
given by
\begin{equation}\label{N}
  \mean{N_f} = \eta\nu\left[\mu|S(\phi)|^2 + (1-\mu)\sinh^2r_2\right],
\end{equation}
with $\nu$ being the transmission of the squeezed mode through the spatial and frequency filters with bandwidth $\Delta\theta$, $\Delta\lambda$, 
$\eta$ the detection efficiency of the setup in the $\Delta\theta$, $\Delta\lambda$ bandwidths, $\mu$ the interferometer internal transmission and
\begin{equation}\label{C,S}
  \begin{array}{rcl}
    S(\phi) &=& \sinh r_1\cosh r_2\,e^{-i\phi} - \cosh r_1\sinh|r_2| \,e^{i\phi} \,.
  \end{array}
\end{equation}
As we work with high gain values, $\sinh r_1\simeq\cosh r_1 \simeq\exp(r_1)/2$ and $\sinh |r_2|\simeq\cosh |r_2| \simeq \exp(|r_2|)/2$ ,  which leads 
to a simplification of Eq.~(\ref{N}),
\begin{equation}
\begin{array}{rcl}
  \mean{N_f} & \simeq & \frac{1}{4}\eta\nu\mu\exp{(2(r_1+|r_2|))} \sin(\phi)^2\\
  &+&\eta\nu(1-\mu)\exp(2|r_2|)/4.
    \end{array}
\end{equation}
Moreover, the mean number of photons inside the interferometer within the $\Delta\lambda$, $\Delta\theta$ bandwidths considered is
\begin{equation}\label{Nins}
  \mean{N(\Delta\lambda,\Delta\theta)} = \nu\sinh^2(r_1)\simeq\nu\exp(2r_1)/4.
\end{equation}
Therefore, 
\begin{equation}\label{N2}
\begin{array}{rcl}
  \mean{N_f} &\simeq& \eta\mu\mean{N(\Delta\lambda,\Delta\theta)} \exp(2|r_2|) \sin(\phi)^2\\
  &&+\eta\nu(1-\mu)\exp(2|r_2|)/4.
\end{array}
\end{equation}
The number of photons in the interferometer $\mean{N(\Delta\lambda,\Delta\theta)}$ is measured with an avalanche photodiode and is therefore known. 
The internal transmission is also known, $\mu=0.97$, as well as the detection efficiency $\eta=0.77$. Thus the only unknown 
parameter in Eq.~(\ref{N2}) is $|r_2|$. The phase independent term $\eta\nu(1-\mu)\exp(2|r_2|)/4$ can be incorporated into a more general 
term, $B$, that also takes into account additional background light due to the non-perfect interference visibility:
\begin{equation}\label{N3}
  \mean{N_f}\simeq \eta\mu\mean{N(\Delta\lambda,\Delta\theta)} \exp(2|r_2|) \sin(\phi)^2+B.
\end{equation}

The parametric gain of the first amplifier $r_1$ and the transmission $\nu$ of the squeezed mode can be subsequently found via 
Eq.~(\ref{Nins}) and the number of photons $\mean{N_2(\Delta\lambda,\Delta\theta)}$ produced by the second crystal in the 
$\Delta\theta$, $\Delta\lambda$ bandwidths, which is also measured: 
\begin{equation}\label{r1}
\begin{array}{ccl}
   r_1&=&\asinh\Big(\sqrt{\dfrac{\mean{N(\Delta\lambda,\Delta\theta)}}{\mean{N_2(\Delta\lambda,\Delta\theta)}}\sinh^2 r_2}\Big ),\\
   \nu&=&\dfrac{\mean{N(\Delta\lambda,\Delta\theta)}}{\sinh^2(r_1)}.
   \end{array}
\end{equation}

\begin{figure}[b]
  \centering
\includegraphics[scale=0.5]{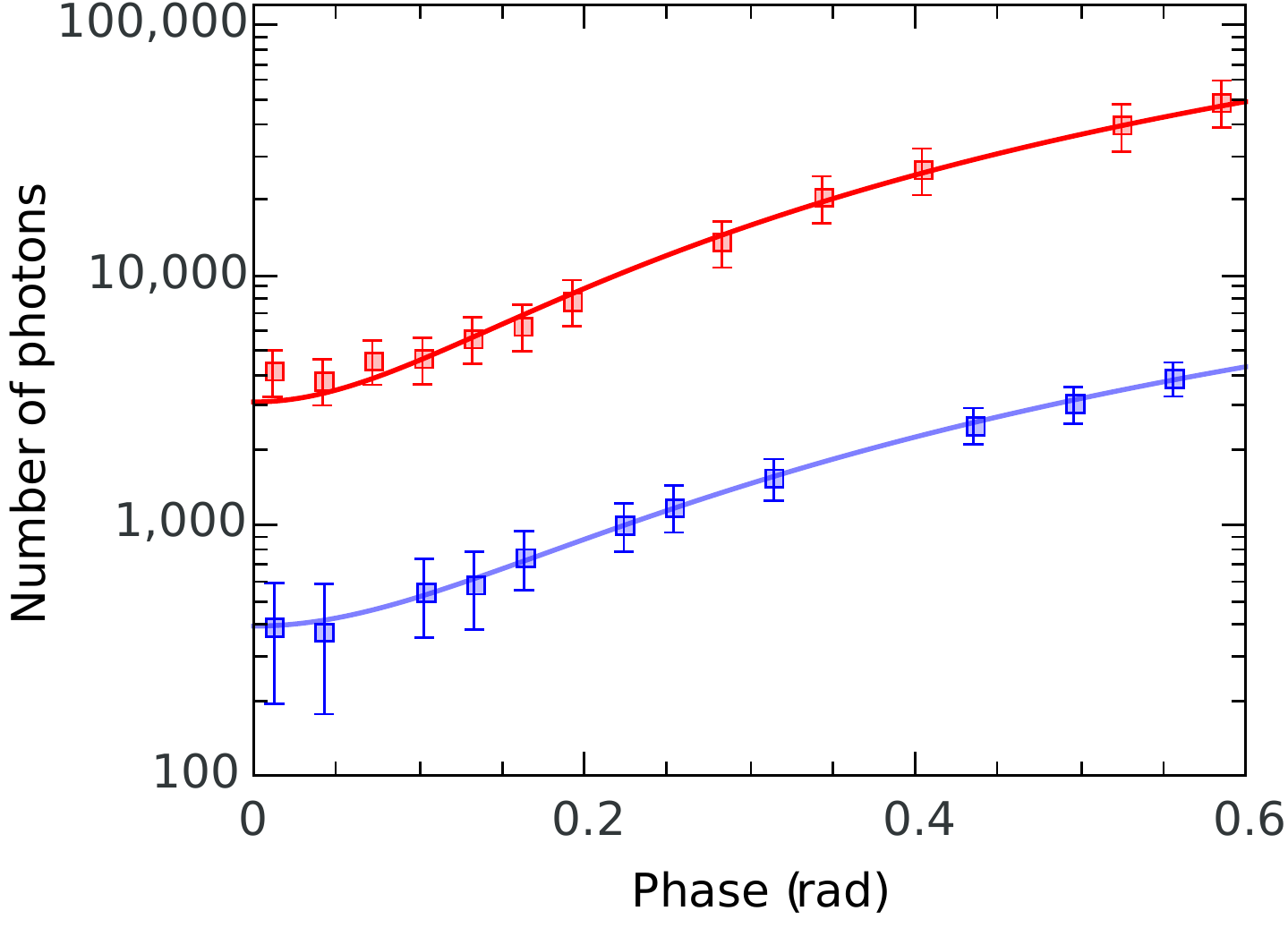}
  \caption{Example of two sets of data fitted by Eq.~(\ref{N2}). Red curve: pump power $140$~mW, $|r_2|=5.2$, $r_1=2.1$ and 
$\nu=0.29$. Blue curve: pump power $90$~mW, $|r_2|=3.9$, $r_1=2.5$ and $\nu=0.13$.}
  \label{fig:Fitting}
\end{figure}

Figure~\ref{fig:Fitting} shows two examples of fit with Eq.~(\ref{N3}) for two datasets. The average pump power was set to $90$~mW for the blue 
curve, 
while it was increased to $140$~mW for the red curve. However the number of photons inside the interferometer $\langle 
N(\Delta\lambda,\Delta\theta)\rangle=4.8$ is the same in both cases as the phase matching of the first crystal was modified in order to keep $\langle 
N(\Delta\lambda,\Delta\theta)\rangle$ constant for the two measurements. The second crystal is aligned such as to maximize the number of photons in 
the $\Delta\lambda$, $\Delta\theta$ windows. Hence the fits with Eq.~(\ref{N3}) yield $|r_2|=3.9$ for the $90$~mW pump power excitation, while the 
larger pump power of $140$~mW  gives a larger gain of $|r_2|=5.2$. Furthermore we measured $\mean{N_2(\Delta\lambda,\Delta\theta)}$, the mean number 
of photons emitted by the second crystal within the bandwidths $\langle\Delta\lambda$,$\Delta\theta\rangle$. Using Eq.~(\ref{r1}), we find  $r_1=2.5$ 
and $\nu=0.13$ for the $90$~mW case, while $r_1=2.1$ and $\nu=0.29$ for the $140$~mW case. It is to be noted that different transmission 
$\nu$ of the 
squeezed mode are found for the two measurements because the mode size changes while changing the phase matching of the first crystal.

\section{Pump pulses post-selection}

The pump beam possesses a relatively good intensity stability with a pulse energy fluctuation around $10$~nJ  for the mean pulse energy of 
$20$~$\mu$J. The corresponding second-order correlation function at zero delay is $g^{(2)}(0)=\frac{\langle I^2 \rangle}{\langle  I \rangle^2} 
\approx 
1.00001$, with $I$ being the measured pump beam intensity. The pump intensity distribution is therefore almost Poissonian ($g^{(2)}(0)\approx 1$). 
However because of the high parametric gains involved in the measurements, these small-scale pump intensity fluctuations are amplified in the PDC 
light. We use an additional BBO crystal BBO$_{pp}$  to enhance these fluctuations and be in the sensitive wavelengths range of the p-i-n diode used. 
The broadband, multimode PDC light generated by BBO$_{pp}$ is then measured by a second p-i-n diode PD2 similar to PD1. A post-processing window is 
set to discard  from the measurement set all excitation pulses outside a chosen interval. Hence, the pump fluctuations are decreased and the mean 
number of photons for a given interferometer position can be measured precisely.

\end{document}